\documentclass[aps,pre,superscriptaddress,twocolumn]{revtex4-1}

\usepackage{graphicx}
\usepackage{dcolumn}
\usepackage{bm}
\usepackage{amsmath}
\usepackage{color}
\usepackage[colorlinks,
linkcolor=blue,
citecolor=blue,
urlcolor=blue]{hyperref}

\begin{document}

\title{Simulating the external magnetic field in short-pulse intense laser-plasma interaction}

\author{K. Jiang}
\affiliation{Graduate School, China Academy of Engineering Physics, Beijing 100088, China}
\affiliation{Center for Advanced Material Diagnostic Technology, and College of Engineering Physics, Shenzhen Technology University, Shenzhen 518118, China}

\author{C. T. Zhou} \email{zcangtao@sztu.edu.cn}
\affiliation{Center for Advanced Material Diagnostic Technology, and College of Engineering Physics, Shenzhen Technology University, Shenzhen 518118, China}
\affiliation{HEDPS, Center for Applied Physics and Technology and School of Physics, Peking University, Beijing 100871, China}

\author{S. Z. Wu}
\affiliation{Center for Advanced Material Diagnostic Technology, and College of Engineering Physics, Shenzhen Technology University, Shenzhen 518118, China}

\author{H. Zhang}
\affiliation{Center for Advanced Material Diagnostic Technology, and College of Engineering Physics, Shenzhen Technology University, Shenzhen 518118, China}

\author{C. N. Wu}
\affiliation{Graduate School, China Academy of Engineering Physics, Beijing 100088, China}
\affiliation{Center for Advanced Material Diagnostic Technology, and College of Engineering Physics, Shenzhen Technology University, Shenzhen 518118, China}

\author{T. Y. Long}
\affiliation{Center for Advanced Material Diagnostic Technology, and College of Engineering Physics, Shenzhen Technology University, Shenzhen 518118, China}
\affiliation{HEDPS, Center for Applied Physics and Technology and School of Physics, Peking University, Beijing 100871, China}

\author{L. Li}
\affiliation{Center for Advanced Material Diagnostic Technology, and College of Engineering Physics, Shenzhen Technology University, Shenzhen 518118, China}
\affiliation{HEDPS, Center for Applied Physics and Technology and School of Physics, Peking University, Beijing 100871, China}

\author{T. W. Huang}
\affiliation{Center for Advanced Material Diagnostic Technology, and College of Engineering Physics, Shenzhen Technology University, Shenzhen 518118, China}

\author{L. B. Ju}
\affiliation{Center for Advanced Material Diagnostic Technology, and College of Engineering Physics, Shenzhen Technology University, Shenzhen 518118, China}

\author{B. Qiao}
\affiliation{HEDPS, Center for Applied Physics and Technology and School of Physics, Peking University, Beijing 100871, China}

\author{M. Y. Yu}
\affiliation{Center for Advanced Material Diagnostic Technology, and College of Engineering Physics, Shenzhen Technology University, Shenzhen 518118, China}

\author{S. P. Zhu} \email{zhu\_shaoping@iapcm.ac.cn}
\affiliation{Graduate School, China Academy of Engineering Physics, Beijing 100088, China}
\affiliation{Institute of Applied Physics and Computational Mathematics, Beijing 100094, China}

\author{S. C. Ruan}
\affiliation{Center for Advanced Material Diagnostic Technology, and College of Engineering Physics, Shenzhen Technology University, Shenzhen 518118, China}

\date{\today}

\begin{abstract}
Imposing an external magnetic field in short-pulse intense laser-plasma interaction is of broad scientific interest in related plasma research areas. We propose a simple method using a virtual current layer by introducing an extra current density term to simulate the external magnetic field, and demonstrate it with three-dimensional particle-in-cell simulations. The field distribution and its evolution in sub-picosecond time scale are obtained. The magnetization process takes a much longer time than that of laser-plasma interaction due to plasma diamagnetism arising from collective response. The long-time evolution of magnetic diffusion and diamagnetic current can be predicted based on a simplified analytic model in combination with simulations.
\end{abstract}

\maketitle

From irradiation of a capacitor-coil target with an intense laser, kilotesla or higher magnetic fields have been generated \cite{Fujioka,Santos,Law,Zhu}. The effect of such strong external magnetic fields (EMFs) in laser-induced high energy density plasmas (HEDPs) has been theoretically and experimentally investigated, such as in the propagation of energetic particles \cite{Hosokai,Kar,Nakamura,Strozzi,Wang,Grandvaux,Sakata,Jiang,Bolanos}, modification of the plasma conductivity \cite{Askaryan,Gong}, magnetic reconnection \cite{Fiksel}, laboratory astrophysics \cite{Mandrini,Honda,Ciardi}, material science \cite{Reitzenstein,Yoneda}, etc. In these studies, particle-in-cell (PIC) simulations played an important role in predicting and explaining the found phenomena.

In most existing PIC simulations involving magnetized plasmas in short-pulse laser-plasma interactions (SLPI), EMFs are introduced by directly assigning matrices of a pre-set magnetic field, and no check is made to ensure that the plasma-vacuum boundary conditions are satisfied. Therefore, a plasma particle is usually set to experience the same $\bf{B}$ field as it does in the vacuum. The interplay between EMF evolution and plasma response is thus ignored.

Since diamagnetism originating from collective response to the EMF is a fundamental property of plasmas exposed to the latter \cite{Chen}, the magnitude and configuration of the latter inside the plasma may not be arbitrary. Therefore, the field felt by plasma particles shall deviate from its vacuum value, which however can not be correctly described by conventional simulations. Recent studies showed that magnetic field penetration into an unmagnetized plasma is on the nanosecond (ns) time scale \cite{Grandvaux, Sakata, Morita}. On the other hand, typical SLPI take place within the scale of picoseconds (ps) or less. Such a large time scale gap suggests that a proper modeling of EMFs is of great necessity to clarify the field distribution and its evolution with sub-picosecond temporal resolution. This is especially important for the EMF applications in various contexts such as nuclear fusion and laboratory astrophysics.

\begin{figure}
\centering
\includegraphics[width=8.6cm]{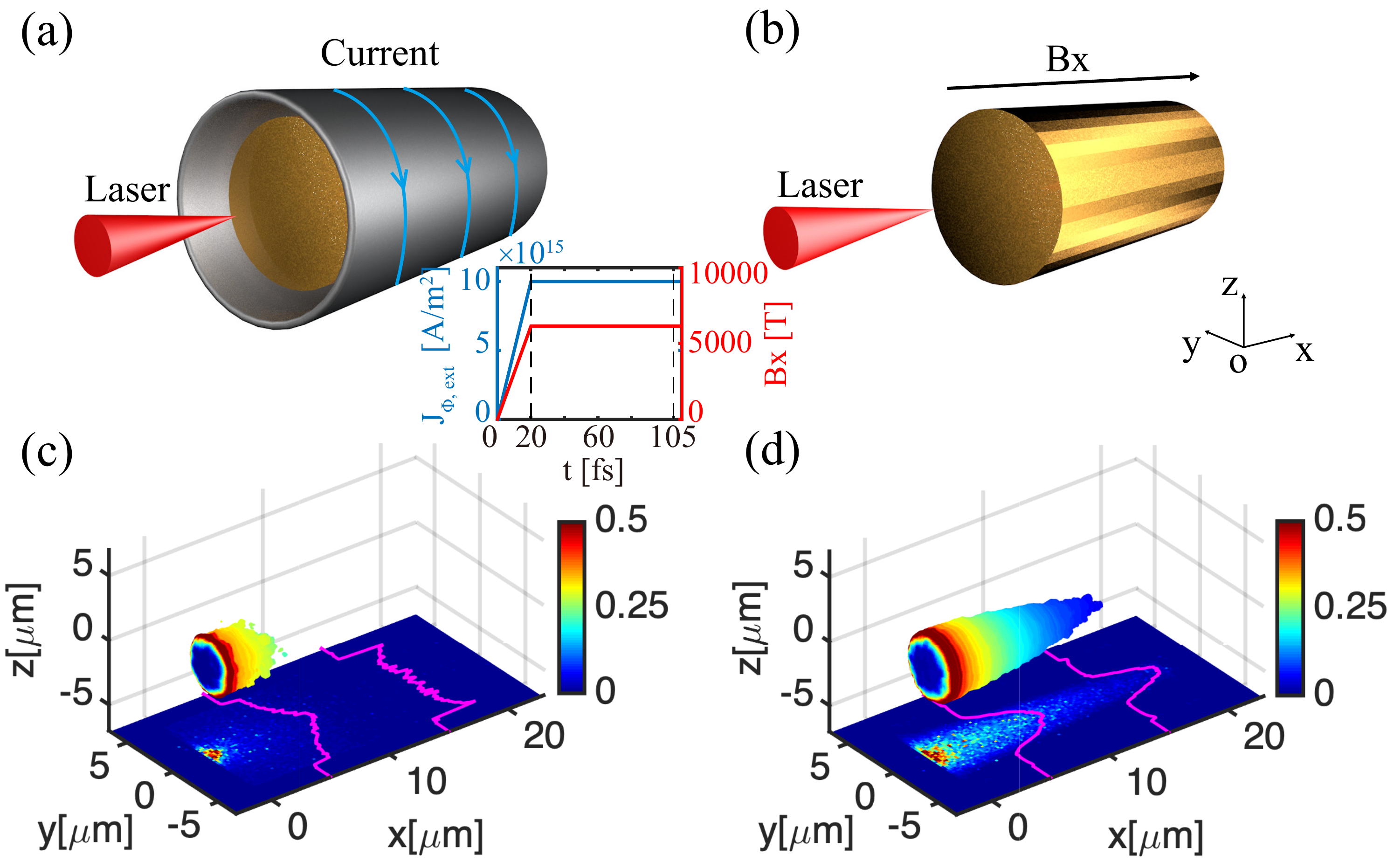}
\caption{(a) Case 1, for the proposed scheme. The grey hollow cylinder with the blue arrows show the virtual current layer and the direction of the current, respectively. The inset in (a) shows the evolution of the current density and the induced magnetic field in the absence of plasma. (b) Case 2 for plasma with imbedded axial magnetic field. Results of the 3D-PIC simulations: the isosurfaces of electron energy density at $t=330$ fs for (c) Case 1 and (d) Case 2. The isovalue is normalized to $1.15$MeV$n_c$ and the color depicts the mean electron energy density on each of the $(y, z)$ planes. The projection of the $z=0$ plane at the bottom shows the spatial electron energy density distributions in arbitrary units, and the magenta curves show the transverse profiles of the electron energy density at $x=5$ and $15$ $\mu$m.\label{fig_1}}
\end{figure}

In this Letter, we propose a simple approach to consider the EMF in short-pulse intense laser-plasma interaction. Unlike previous models, a virtual external current layer is introduced as the EMF source. It resembles the capacitor-coil target configuration for generating kilotesla magnetic fields in experiments \cite{Fujioka, Santos, Law, Zhu}. As indicated in Fig. \ref{fig_1}(a), a cylindrical plasma is located in the hollow of a concentric azimuthal current. In this way, we can investigate the EMF evolution and its coupling with plasmas self-consistently, resulting in resolving the field distribution and its evolution in sub-picosecond. 

Our scheme is illustrated by three-dimensional (3D) particle-in-cell (PIC) code EPOCH3D \cite{Arber} with appropriate modifications. The current density after a standard PIC loop is updated by introducing an extra term which can be tuned in order to get the desired EMF. The electric and magnetic fields are then updated by the field solver using a Yee staggered Finite-Difference Time-Domain scheme \cite{Arber}. The simulation configuration (Case 1) is visualized in Fig. \ref{fig_1} (a). The Cu$^{2+}$ plasma is $R=5$ $\mu$m in radius and $L=20$ $\mu$m in length, with electron and ions densities $n_{e0}=20n_c$ and $n_{i0}=10n_c$, respectively, where $n_c\sim 1.1\times 10^{21}$ cm$^{-3}/\lambda_L^2$ is the critical density and $\lambda_L$ the incident laser wavelength in units of $\mu$m. The initial electron temperature is $T_{e0}=100$ eV, and the ion temperature $T_{i0}=10$ eV. The cylindrical current layer is 0.5 $\mu$m thick and 29 $\mu$m long. The current density rises linearly from 0 to $1\times 10^{16}$ A/m$^2$ in 20 fs and remains constant thereafter. Thus, without the plasma the current layer would generate a uniform axial magnetic field of 6240 T in the cylindrical hollow, as shown in the inset of Fig. \ref{fig_1}. A $y$-polarized Gaussian laser pulse of $\lambda_L = 1.06$ $\mu$m, peak intensity $3 \times 10^{19}$ W/cm$^2$, FWHM spotsize $r_0=3.3$ $\mu$m, and duration 420 fs, is normally incident from the left boundary (at $x=-3$ $\mu$m) 85 fs after the current in the layer attains its maximum. The laser has a flat-top temporal profile, after a 17.5 fs rising time. The simulation box is of size $L_x\times L_y\times L_z = 26\times 14\times 14$ $\mu$m$^3$, with $1143\times 795\times 795$ grids and 4 macro electrons and 2 macro ions per cell. For comparison, Fig. \ref{fig_1}(b) shows a conventional simulation configuration (Case 2), which starts with a fully magnetized plasma with uniform axial EMF $B_{0x,\mathrm{ext}}=6240$ T, using the standard ``fields block'' in EPOCH3D \cite{Arber}. The other parameters are also the same as these in Case 1.

\begin{figure}
\centering
\includegraphics[width=8.6cm]{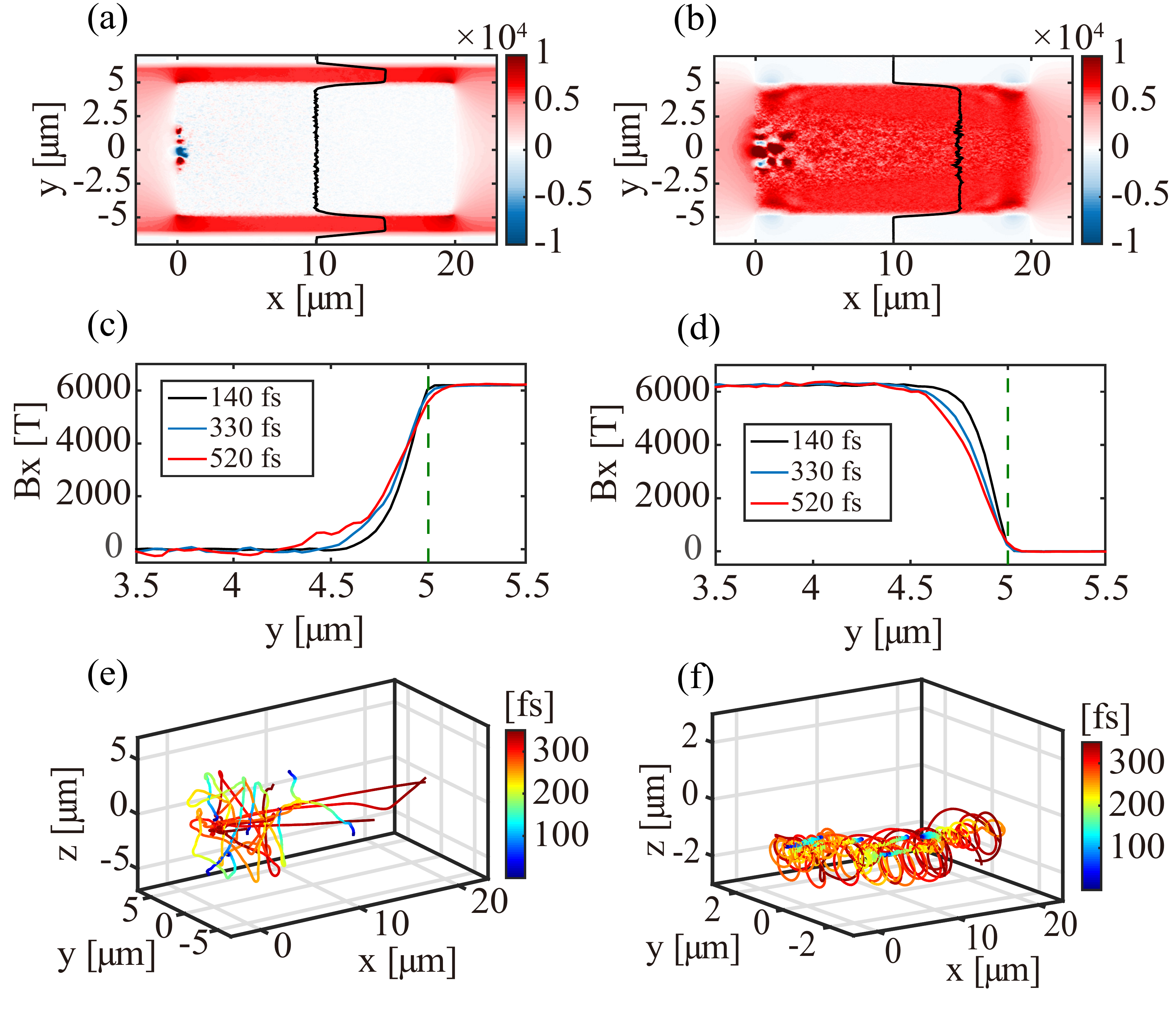}
\caption{Results of the 3D-PIC simulations. Magnetic field $B_x$ profiles on the $z=0$ plane at $t=330$ fs for (a) Case 1 and (b) Case 2. The black curves in (a) and (b) show the transverse profiles of $B_x$ at $x=10$ $\mu$m. Enlargements of the $B_x$ transverse profiles at $x=10$ $\mu$m for (c) Case 1 and (d) Case 2. The green dash lines mark the plasma-vacuum boundary. The trajectories of 7 typical electrons in (e) Case 1 and (f) Case 2. \label{fig_2} }
\end{figure}

Figures \ref{fig_1}(c) and (d) show the energy density of the intense laser-produced energetic electrons at $t=330$ fs for the Cases 1 and 2, respectively. Obvious difference between them can be found. Fig. \ref{fig_1}(c) for Case 1 shows that the energetic electrons remain mainly near the target front surface, and the electron energy density (EED) profile has a transverse FWHM close to the laser spotsize. The EED gradually broadens and becomes nearly homogeneous inside the plasma, as indicated by the magenta curves at $x=5$ and 15 $\mu$m. At $x=5$ $\mu$m, the EED at the center (within the laser spot area) accounts for $20.8\%$ of the total EED on the $(y,z)$ plane, but drops to $11.7\%$ at $x=15$ $\mu$m since the electrons diverge as they propagate forward. On the other hand, Fig. \ref{fig_1}(d) for Case 2 shows that the laser-accelerated electrons are well guided by the embedded magnetic field as they propagate deep into the plasma. The transverse EED profile also remains Gaussian-like, with FWHM around 3.4 $\mu$m. At $x=5$ $\mu$m, the EED within the laser spot area accounts for $64.3\%$ of the total, and it only drops to $60.5\%$ at $x=15$ $\mu$m. The trajectories of 7 typical electrons for Cases 1 and 2 in Figs. \ref{fig_2}(e) and (f), respectively, further confirm these observations.

As expected, Figs. \ref{fig_2} (a) and (b) show that the magnetic field profiles are quite different for the two cases. In Case 1, within the plasma there is effectively no magnetic field. The field drops from the vacuum value (6240 T) to about null within a boundary layer of thickness less than 1 $\mu$m. In fact, the panels (a) and (c) indicate that the magnetic field in the bulk plasma remains almost everywhere null throughout the simulation. The dynamics of the laser-accelerated electrons is similar to that in unmagnetized dense plasma \cite{Pukhov}, as can be seen in the panel (e). On the other hand, in Case 2 the axial magnetic field remains uniform in the plasma bulk. The gyroradius of the affected electrons is roughly $r_e\sim 0.5$ $\mu$m.
Since $r_e$ is much smaller than the plasma radius, the electrons are well guided by the EMF, as can be seen in the panel (f), as well as in Fig. \ref{fig_1}(d). The results here, especially the black curves for $B_x$ in the panels (a) and (b), show that the magnetic field induced by the laser-driven electron beam does not significantly modify the EMF, whose distribution is however very different in the two cases.

It is interesting to note that the EMF for both cases are identical in vacuum. However, the physically expected magnetic guiding effect which appeared in Case 2, did not take place in Case 1. The significant differences in the field distribution and electron dynamics can be attributed to the simulation methods of modeling the EMF. In Case 1 (using a virtual current layer), the magnetic field evolution and its interplay with plasmas can be clearly observed. A further analysis will be followed. In Case 2 (using a pre-set EMF), however, plasma is just a collection of single particles moving under the EMF, and collective plasma response is thus absent.

\begin{figure}
\centering
\includegraphics[width=8.6cm]{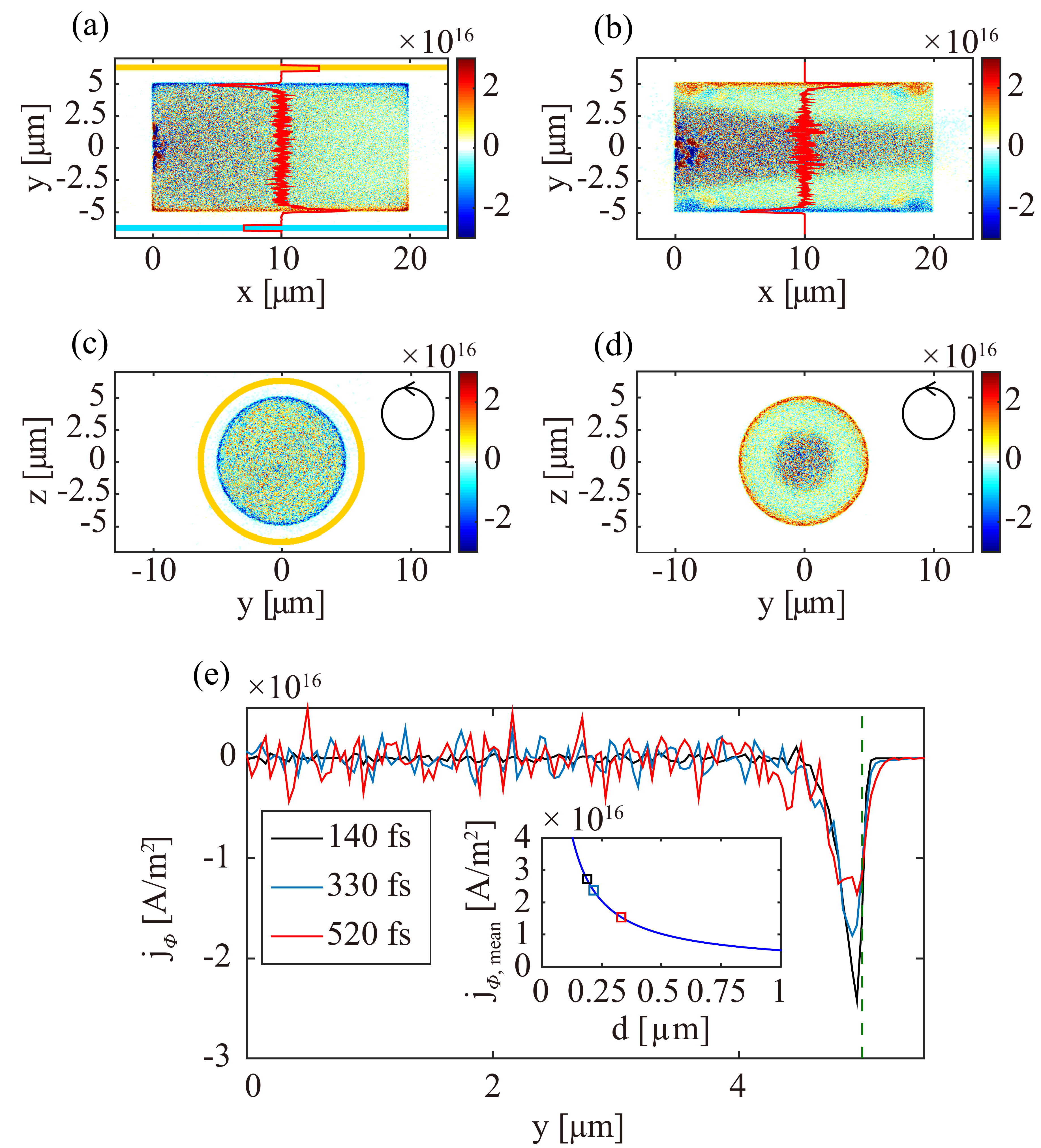}
\caption{Distribution of the azimuthal current density $j_\phi$ in A/m$^2$ on the $z=0$ plane at $t=330$ fs for (a) Case 1 and (b) Case 2. The red curves show the transverse profile of $j_\phi$ at $x=10$ $\mu$m. The corresponding cross-sectional ($y,z$) plane views are given in (c) and (d), and the black arrows on the upper right corners indicate the positive $j_\phi$ direction. (e) Enlargement of a $y>0$ section of the $j_\phi$ versus $y$ profile in (c) for the Case 1, in the $x=10$ $\mu$m plane at $t=140, 330$, and $520$ fs. The vertical green dash line marks the plasma-vacuum boundary. The inset shows the dependence of $j_{\phi, {\rm mean}}$ on the thickness $d$ of the diamagnetic current layer from the analytical result based on Eq. (\ref{eq1}), and the squares mark the points corresponding to the different times. \label{fig_3}}
\end{figure}

An azimuthal current layer can be observed at the plasma-vacuum boundary for both cases, but with opposite directions, as shown in Figs. \ref{fig_3} (a)-(d). In Case 1, the current layer is due to the diamagnetic current formed in the plasma edge as a result of response of the plasma particles to the EMF, and it acts to prevent penetration of the magnetic field into the bulk plasma. The result here agrees well with that from the plasma theory \cite{Chen} and experiments \cite{Hallock}. Comparison of Fig. \ref{fig_3}(e) with Fig. \ref{fig_2}(c) suggests that the thickness $d$ of the diamagnetic current layer matches that of the magnetic field penetration into the plasma. Moreover, $d$ increases, but the maximum value of the diamagnetic current density decreases, slowly with time. Assuming that the total magnetic field at the center of the plasma remains null, we can obtain an analytical expression for the mean value of the diamagnetic current density based on Biot-Savart law \cite{Jackson}
\begin{equation}
j_{\phi,{\rm mean}}=\frac{B_{0x,{\rm ext}}}{\mu_0L\ln\left[ \frac{\sqrt{L^2+R^2}+R}{\sqrt{L^2+(R-d)^2}+R-d}\right], } \label{eq1}
\end{equation}
which is shown in the inset of Fig. \ref{fig_3}(e). These results suggest that the virtual current approach should be useful for simulating magnetization of dense plasmas.

\begin{figure}
\centering
\includegraphics[width=8.6cm]{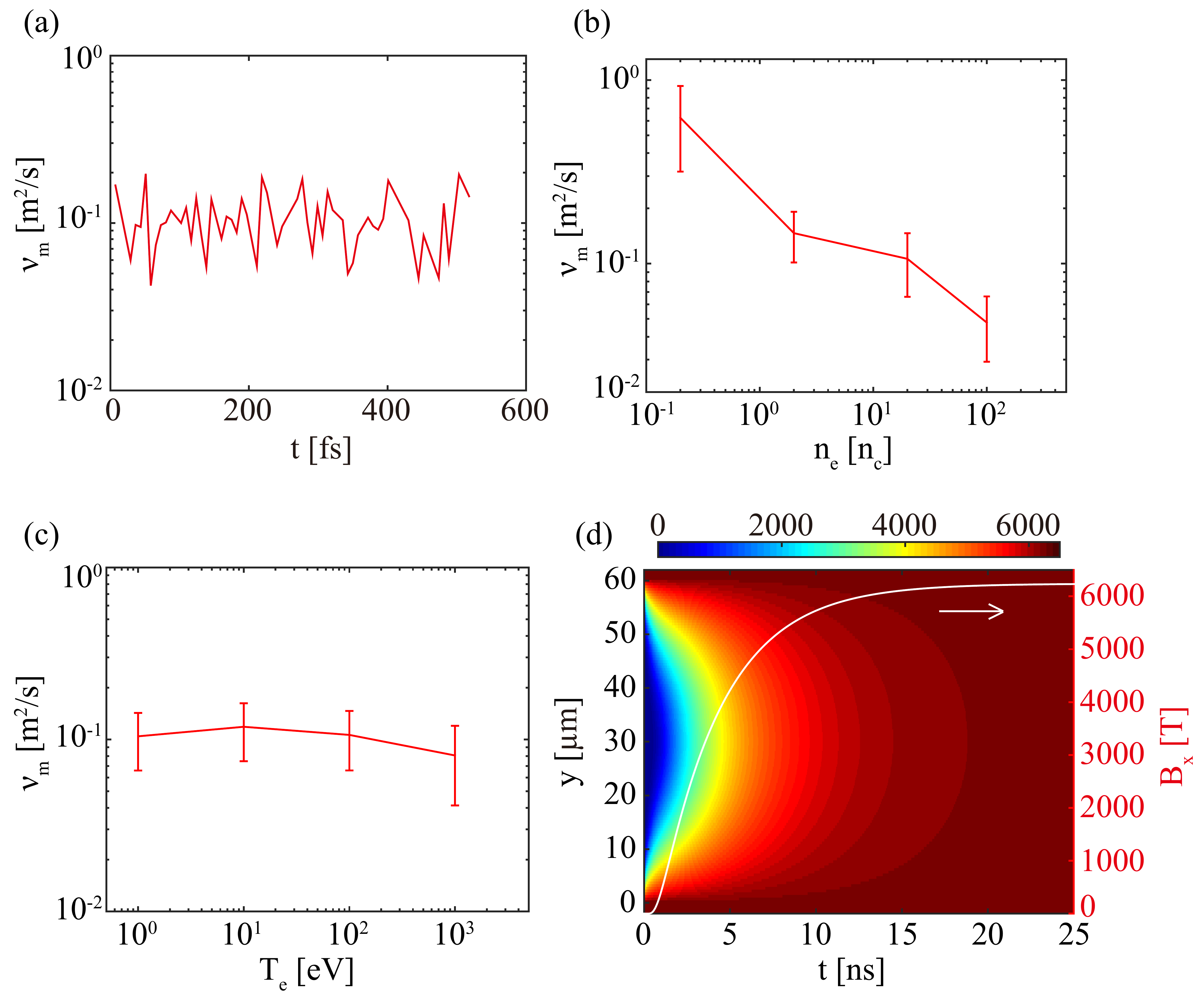}
\caption{(a) Evolution of the magnetic diffusion coefficient $\nu_m$ as calculated from the simulation results for $n_e=20n_c$ and $T_e=100$ eV. $\nu_m$ as a function of (b) the electron density for $T_e=100$ eV and (c) the electron temperature for $n_e=20 n_c$. (d) Evolution of the magnetic field obtained by numerically solving Eqs. (\ref{eq2})-(\ref{eq4}). The white curve shows the evolution of $B_x$ at the central axis $y=30$ $\mu$m.\label{fig_4}}
\end{figure}

It may thus be of interest to investigate in more detail the characteristics of the magnetic diffusion in the plasma. We therefore perform new simulation runs where the irradiating intense laser is switched off to preclude the effect of the laser pulse on the magnetic diffusion process. The evolution of magnetic fields in plasma can roughly be described by the diffusion equation \cite{Chen} $\frac{\partial{\bf B}}{\partial t}=\nu_m\nabla^2{\bf B}$, with $\nu_m$ representing the magnetic diffusion coefficient. For simplicity, we consider a quasi one-dimensional geometry, such that the evolution of the diffusion process can be approximated by
\begin{equation}
\frac{\partial B_x}{\partial t}=\nu_m\frac{\partial^2 B_x}{\partial y^2}, \label{eq2}
\end{equation}
with the boundary conditions
\begin{equation}
B_x(y=0)=B_0 \quad {\rm and} \quad B_x(y=l)=B_0,  \label{eq3}
\end{equation}
and initial conditions
\begin{equation}
B_x(y,t=0)=\left\{\begin{matrix}
B_0, \;y =  0 \; {\rm or}\; y = l \\
0, \;0<y<l.  \label{eq4}
\end{matrix}\right.
\end{equation}

Although the general solution of the above system is readily available, for our purpose it is more convenient to solve Eqs. (\ref{eq2})-(\ref{eq4}) numerically using a finite difference method. The instantaneous value of the magnetic diffusion coefficient $\nu_m$ can be numerically calculated by carefully following the penetration, say a leading isosurface, of the diffusing magnetic field in the PIC simulation. The resulting evolution of $\nu_m$ for the Case 1 is given in Fig. \ref{fig_4}(a). We see that $\nu_m$ fluctuates on the fs timescale, but on longer timescales it has a well-defined mean value, and that the mean field diffusion coefficient is almost time independent. Accordingly, the magnetic diffusion coefficient $\nu_m$ so obtained can be used to estimate the behavior of the diffusion of EMFs into plasmas. For the Case 1 (for $n_e=20n_c$ and $T_e=100$ eV), we found $\nu_m \sim 0.1$ m/s$^2$. We have also varied the initial electron density (from $0.2 n_c$ to 100$n_c$) and temperature (from 1.0 eV to 1000 eV). Similar fluctuation behavior is found (not shown) in all cases. Figs. \ref{fig_4}(b) and (c) show $\nu_m$ as function of the electron density and temperature, respectively. We see that $\nu_m$ decreases with the increase of $n_e$, but is only weakly dependent on $T_e$. This result is quite different from that of collision dominated magnetic diffusion, where according to the Spitzer-Harm theory $\nu_m$ would depend mainly on the electron temperature \cite{Morita}. Such a discrepancy will be investigated in a future work. Fig. \ref{fig_4}(d) shows the evolution of the magnetic field in a typical HEDP case of short-pulse laser-plasma interaction as obtained from the numerical solution of Eq. (\ref{eq2})-(\ref{eq4}), with the plasma and external magnetic field parameters the same as that in Case 1, except that the transverse size $l=60$ $\mu$m. For the calculation we have used $\nu_m=0.1$ m/s$^2$. One can see that full magnetization of the plasma occurs on the ns to 10 ns time scale, consistent with that from hydrodynamic simulations and some existing works \cite{Grandvaux,Sakata,Morita}. From this point of view, the model using the external field in Case 2 can also be considered as a long-time stationary situation of Case 1, where the plasma is fully magnetized and thus no diamagnetism is observed.

\begin{figure}
\centering
\includegraphics[width=8.6cm]{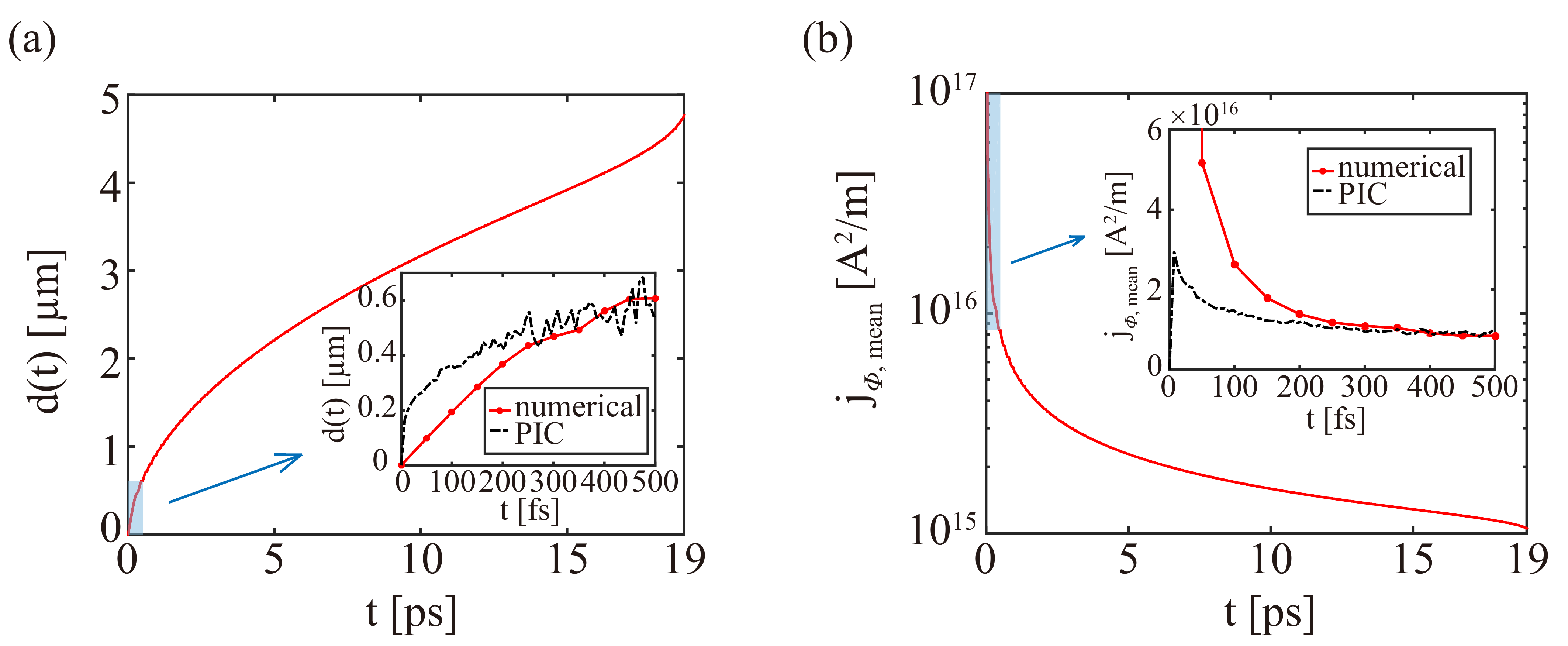}
\caption{Evolution of (a) the diamagnetic current thickness $d(t)$ and (b) its mean value $j_{\phi,{\rm mean}}$ numerically calculated for $\nu_m=0.1$ m$^2$/s. The insets show enlargements of the initial stage (light-blue areas) including PIC simulation results (black dash curves).\label{fig_5} }
\end{figure}

As discussed above, the diamagnetic current is strongly correlated to the magnetic field diffusion. In Case 1, by following the field evolution, we can deduce the evolution of the diamagnetic current. Without loss of generality, the thickness $d(t)$ of the diamagnetic current layer is defined as the distance in which the magnetic field drops to 10\% of its initial value in the vacuum region. Fig. \ref{fig_5}(a) shows the evolution of $d(t)$ as obtained by solving Eqs. (\ref{eq2})-(\ref{eq4}) numerically, with the boundary and initial conditions the same as in the PIC simulations for Case 1. It can be seen that $d(t)$ increases with time, as to be expected, which suggests that it may be used to represent the spatial extent of magnetization process, as a supplementary to the ratio of cyclotron frequency and plasma frequency commonly used in laboratory astrophysics \cite{Yao}. From Eq. (\ref{eq1}), one can then readily obtain the evolution of the diamagnetic current density, which is shown in Fig. \ref{fig_5}(b). We see that $j_{\phi,\rm{mean}}$ initially drops rapidly and then decays in a rather slow rate. It is also noted that both $d(t)$ and $j_{\phi,\rm{mean}}$ and their behavior obtained numerically agree well with that of the PIC simulation for the entire simulation time (as shown in the insets in Figs. \ref{fig_5}(a) and (b)), which in a sense justifies our highly simplified analytical model. The discrepancy for $t<200$ fs can be attributed to the uncertainty of the numerical boundary, which however does not affect the later development.

In conclusion, we propose a simple method using a virtual current layer to simulate the external magnetic field in short-pulse laser-plasma interaction. 3D PIC simulations show that the field distribution as well as its evolution in sub-picosecond time scale can be obtained. The typical magnetization process take places on ns to 10 ns time scale due to plasma diamagnetism, long after laser-plasma interaction has ended. The long-time evolution of the magnetic diffusion and the diamagnetic current is to be predicted by combining a simplified analytical model and simulations. Our scheme can be easily extended to arbitrary magnetic field configuration, and should be useful in HEDP studies, especially nuclear fusion, laboratory astrophysics, etc.\\

This work is supported by the National Key R\&D Program (Grant No. 2016YFA0401100), the National Natural Science Foundation of China (Grant Nos. 11575031, 11705120, U1630246 and 11875092),  Science Challenge Project (Grant No. TZ2016005), the Natural Science Foundation of Top Talent of SZTU (grant no. 2019010801001).  K. J. would like to thank R. Li, R. X. Bai, and D. B. Zou for useful discussions and help.

K. Jiang and S. Z. Wu contributed equally to this work.

\end{document}